\documentclass[aps,twocolumn,superscriptaddress,floatfix,longbibliography]{revtex4-1}
\usepackage{graphicx,epsf,amsmath,amssymb,wasysym,verbatim,color,graphicx,cleveref}
\usepackage{multirow}
\usepackage{ulem}

\begin{document}
\title{Identifying influential subpopulations in metapopulation epidemic models using message-passing theory}
\author{Jeehye Choi}
\affiliation{Research Institute for Nanoscale Science and Technology, 
	Chungbuk National University, Cheongju, Chungbuk 28644, Korea}
\author{Byungjoon Min}
\email{bmin@cbnu.ac.kr}
\affiliation{Research Institute for Nanoscale Science and Technology, 
	Chungbuk National University, Cheongju, Chungbuk 28644, Korea}
\affiliation{Department of Physics, Chungbuk National University, Cheongju, Chungbuk 28644, Korea}
\date{\today}

\begin{abstract}
Identifying influential subpopulations in metapopulation epidemic models has far-reaching potential 
implications for surveillance and intervention policies of a global pandemic. 
However, there is a lack of methods to determine influential nodes in metapopulation 
models based on a rigorous mathematical background. In this study, we derive the 
message-passing theory for metapopulation modeling and propose a method to determine 
influential spreaders. Based on our analysis, we identify the most dangerous 
city as a potential seed of a pandemic when applied to real-world data. Moreover, we particularly 
assess the relative importance of various sources of heterogeneity at the 
subpopulation level, e.g., the number of connections and mobility patterns, to determine 
properties of spreading processes. We validate our theory with extensive 
numerical simulations on empirical and synthetic networks considering various mobility and 
transmission probabilities. We confirm that our theory can accurately predict
influential subpopulations in metapopulation models.
\end{abstract}
\maketitle

\section{Introduction}

The spread of infectious diseases has significantly increased worldwide owing to the 
development of mobility technologies \cite{Anderson2004,Hollingsworth2006,Colizza2006,
Jones2008,Arenas2020,Kraemer2020}. For instance, we have been threatened 
by infectious diseases such as SARS in 2002, influenza H1N1 in 2009, West Africa Ebola 
outbreak in 2013–2016, and COVID-19, which is currently impacting the entire globe 
\cite{Anderson2004,Hollingsworth2006,Maier2020}. To understand and control 
the spread of infectious diseases, the quantitative modeling of spreading processes is 
required \cite{May1991,Newman2002,Pastor-satorras2015,Maier2020,Arenas2020,Siegenfeld2020,Murphy2021}.
Several traditional studies on epidemics have assumed well-mixed populations with  
a compartmental model \cite{May1991} or a networked system where a node corresponds to a single 
individual \cite{Pastor-satorras2001,Newman2002,Castellano2009,Pastor-satorras2015}. 
However, many real-world systems in a global scale are better described by metapopulation 
models \cite{Colizza2006,Colizza2007,Colizza_natphys2007,Colizza2008,Masuda2010,Apolloni2014,Panos2018}.
In metapopulation modeling, each node corresponds to a subpopulation that 
is generally assumed to be internally well-mixed
(e.g., a city), while a sparse network of interconnections exists between nodes.
The hierarchical structure in metapopulation models provides a natural substrate to describe 
realistically spreading phenomena occurring over multiple scales \cite{Colizza2007,Durrett2014,Gracia2014}. 
On the microscopic scale, individuals within each subpopulation are assumed to be well-mixed; i.e., an 
infectious individual infects susceptible individuals in the same subpopulation with a homogeneous 
probability. On the macroscopic scale, individuals move from one subpopulation to another
through interconnections between them. The combination of contagion processes in the microscopic scale 
and migration processes in the macroscopic scale helps metapopulation models exhibit more realistic 
epidemic processes \cite{Colizza2007,Costa2020,Castioni2021}.

Let us consider a scenario of epidemic propagation on metapopulation models. Initially, a few 
infected individuals in a single subpopulation start to spread a disease within the 
subpopulation. At the same time, the infected individuals move to another connected 
subpopulation by chance and, then, spread the disease in the moved subpopulation. 
Over a long time, the epidemic may spread globally or end in a small subpopulation. 
The final consequence of epidemic transmission is significantly influenced by the 
location and characteristics of the subpopulation, from which the epidemic initially 
begins \cite{Colizza2006,Nicolaides2012,Hasegawa2016,Gardenes2018}. As each subpopulation affect 
the spreading process heterogeneously, it is necessary to accurately predict influential 
subpopulations for effective prediction and control of epidemics \cite{Tanaka2014,Matsuki2019}.
In recent times, several methods have been attempted to determine important nodes in a 
networked system \cite{Freeman1977,Albert2000,Dorogovtsev2006,
Kitsak2010,Morone2015,Radicchi2016,Lokhov2017,Iannelli2018,Pei2020,Li2021Identifying}. 
However, there is no guarantee that the method of determining important nodes in a 
simple network is applicable because the metapopulation network includes hierarchical 
structures \cite{Colizza2007,Gardenes2018,Soriano2020,Cota2021}. 
Moreover, there is still a lack of centrality to determine influential subpopulations 
in a metapopulation epidemic model based on a concrete mathematical background.

We focus on predicting, given a seed initiator in subpopulation, what is the probability 
that an extensive epidemic outbreak occurs and what is its expected size. 
We regard that the higher the probability and size of causing global epidemics, the
higher the influential spreading nodes (e.g., cities) are. 
We propose a systematic way to identify the most influential subpopulations
in a metapopulation model based on a message-passing (MP) approach \cite{Karrer2014,Morone2015,Min2020}. 
With regard to advantages, the MP approach can be applied to a single network
rather than an ensemble of networks, include a node-level analysis, and is based on 
a solid theoretical background \cite{Karrer2014}. Based on our analysis, we propose a method to determine 
the most dangerous city as a potential seed of a pandemic when applied to real-world data.
Moreover, we particularly assess the relative importance of various sources of heterogeneity 
at the subpopulation level, such as population abundance, 
number of connections, and mobility patterns in determining overall properties of the 
spreading process. We validate our theory with extensive numerical simulations on synthetic and 
empirical networks considering various scenarios.

\section{Metapopulation Epidemic Model}

\begin{figure}\centering
\includegraphics[width=1\linewidth]{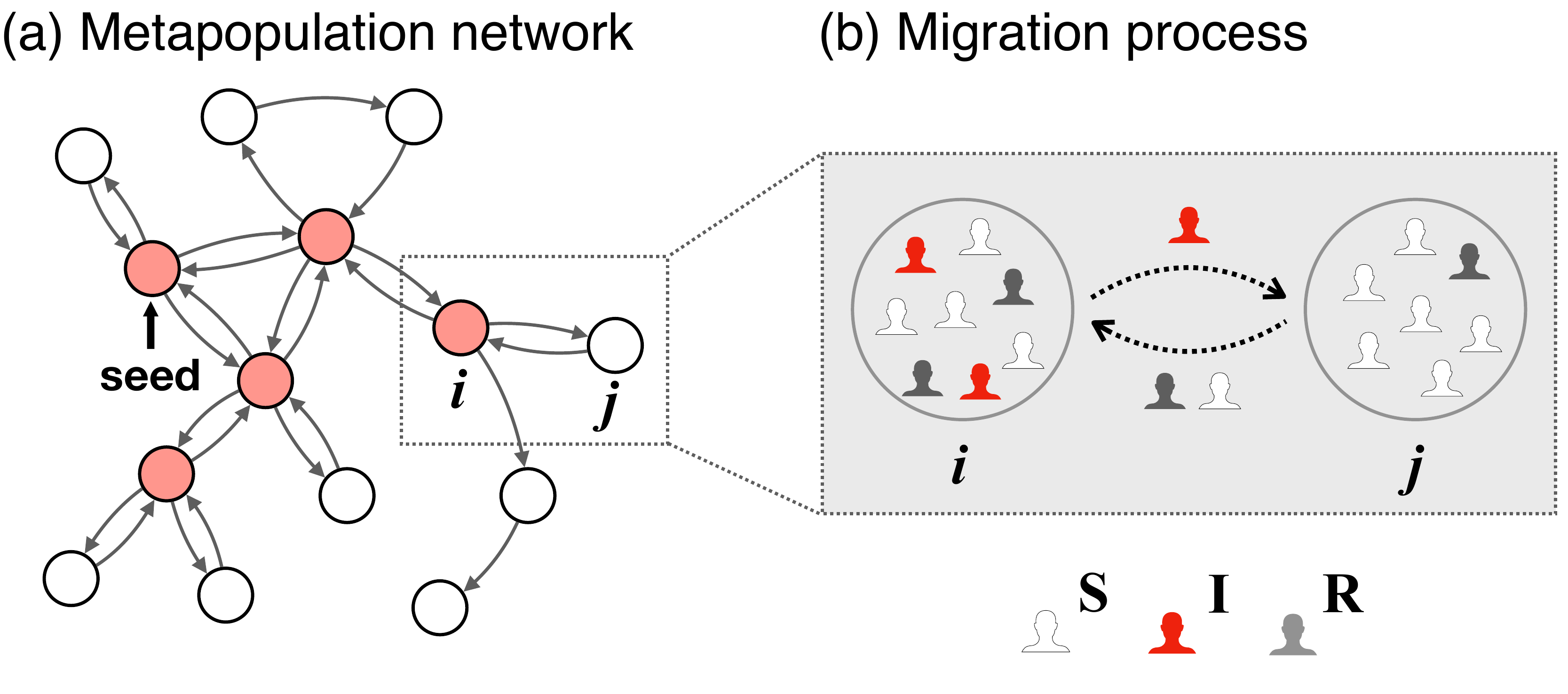}
\caption{
Schematic of a metapopulation epidemic model. Each node corresponds to a 
subpopulation that is internally well-mixed (e.g., a city) while interconnections 
of a sparse metapopulation network exist between nodes. 
The interconnections between subpopulations are in general directed and weighted.
A disease starts to spread from a single seed on a metapopulation network and 
infectious individuals spread the disease to susceptible individuals in the same 
subpopulation. The propagation of the disease to neighboring subpopulations is 
caused by the migration of infectious individuals from a subpopulation in the 
epidemic phase.
}
\label{model}
\end{figure}

Let us consider a network of $N$ subpopulations with interconnections among them. 
The population of subpopulation $i$ is denoted as $n_i$. Individuals in a subpopulation 
are well-mixed and homogeneously interact with those who stay in the same 
subpopulation. In other words, dynamics within each subpopulation can be well 
described based on the naive mean-field approximation. In addition to the dynamics 
within each subpopulation, individuals in each subpopulation can travel to an adjacent 
subpopulation with a certain migration rate. Such hierarchical structure in metapopulation 
models can exhibit realistic modeling for epidemics in a global scale [Fig.~1(a)].

In addition, we consider the spread of a disease described by 
the susceptible-infected-recovered (SIR) model. 
Each individual can possess any one of the following three states: susceptible, infected 
(or infectious), and recovered (or removed). Each susceptible individual becomes infected with 
at infection rate $\beta$ when they contact infected individuals. Infected individuals 
autonomously recover at recovery rate $\gamma$. We define the basic reproduction number
as $R=\beta/\gamma$. These infection and recovery processes occur in each subpopulation according 
to the conventional SIR model. The spread of the disease between
connected subpopulations is mediated through the migration process. When infectious
individuals move to a connected subpopulation, the disease can spread within the subpopulation [Fig.~1(b)].

The consequence of disease spread is dependent on the subpopulation where the disease 
initially began to spread. The more the influential spreading subpopulations are, 
the higher the probability of causing global epidemics and the larger the affected 
size (of the subpopulation). To assess the influence $G_i$ of seed location, we measure 
the number of subpopulations that will eventually be part of the epidemic phase when 
the epidemic starts at subpopulation $i$; i.e., the influence of subpopulation is the 
average fraction $G_i$ of subpopulations (nodes) in the epidemic phase starting 
at node $i$. First, we call that subpopulation $i$ is in the epidemic phase 
when the accumulated number of infections $f_i$ in subpopulation $i$ exceeds 
threshold $\theta n_i$; in this study, we use $\theta=0.01$. 
Then, we define $G_i$ as the fraction of subpopulations that are in the epidemic
phase in a steady state: 
\begin{align}
G_i=\frac{1}{N} \sum_{j=1}^{N} H \left(f_j - \theta  n_j \right), 
\end{align}
where $N$ is the total number of nodes and $H(x)$ is the Heaviside step function 
defined as $1$ for $x>0$ and $0$ for $x \le 0$.
When a significant fraction of subpopulations are in the epidemic phase in the steady 
state, we call that global epidemic occurs; i.e., a global epidemic has occurred with
seed $i$ when $G_i \ge \theta_G$ where $\theta_G=0.01$ in this study.

We use the following individual based simulations to numerically compute the influence $G_i$,
probability $P_i$, and size $S_i$ of epidemics with seed $i$. 
Initially, there is a single infected seed in subpopulation $i$. 
At each discrete time step representing time scale $\tau$, the susceptible individual 
in subpopulation $i$ becomes infected with probability $1-(1-\beta \tau/n_i)^{I_i}$ \cite{Colizza2008}, 
where $n_i$ is the total number of individuals in node $i$ and $I_i$ is the number of infected 
individuals in node $i$. 
The infected individual autonomously recovers 
with probability $\gamma \tau$. In our simulation, we use $\tau=1$.
In addition, each individual in node $i$ attempts to move to a connected subpopulation $j$
with a migration rate. Note that the migration processes in general occur on a weighted and 
directed metapopulation network.
These processes proceed until there are no infected individuals. We repeat 
the procedures for different seeds $i$. We compute the following three measurements: 
i) probability $P_i$ of epidemic outbreak when the epidemic is initiated from 
subpopulation $i$ (origin), ii) the fraction $S_i$ of infected nodes when global epidemic occurs
with $i$, and iii) the fraction $G_i$ of infected nodes in a metapopulation network with epidemics 
starting at node $i$.



\section{Message-passing theory on a metapopulation model}


In this section, we derive MP equations to predict the 
influence of each node in the metapopulation epidemic model. 
In the metapopulation modeling, the probability of disease spreading between connected 
subpopulations is strongly influenced by the total number $m_{ij}$ of infectious movers 
from $i$ to $j$. The probability distribution of $m_{ij}$ can be estimated by 
Binomial distribution such that $\binom{N^{r}_i}{m_{ij}} \mu_{ij}^{m_{ij}} (1-\mu_{ij})^{N^{r}_i-m_{ij}}$,
where $N^r_i$ is the total number of recovered agents of node $i$ in the steady state
which are thus once infectious, and $\mu_{ij}$ is the probability that an infectious  
agent moves from $i$ to $j$ while it is infectious. In the limit $N^r_i \gg 1$, $\mu_{ij} \ll 1$,
and $N^r_i \mu_{ij}$ is a constant $\lambda_{ij}$, the probability distribution of the number of infectious 
movers can be estimated by a Poisson distribution as \cite{Van1992}
\begin{align}
Q(m_{ij}) = \frac{e^{-\lambda_{ij}}  \lambda_{ij}^{m_{ij}} }{m_{ij}!},
\end{align}
where $\lambda_{ij}$ represents the average number of infectious movers from $i$ to $j$.
Since the Poisson distribution $Q(m_{ij})$ is determined by solely $\lambda_{ij}$, 
from now on we mostly use $\lambda_{ij}$ rather than $\mu_{ij}$ and $N^r_i$.
How $\lambda_{ij}$ is given depends on the model details, which is dealt 
in each scenario later.

We next derive the probability $q_{ij}$ of a connected subpopulation $j$ being eventually 
in the epidemic phase caused by an infectious mover originating from subpopulation $i$. 
In other words, $q_{ij}$ represents the probability that node $j$ will be in the epidemic 
phase based on infectious agents moving from node $i$. If infectious agents arrive at $j$, they 
play a similar role as seed agents in subpopulation $j$. The probability of epidemic outbreaks 
with $m$ seeds can be estimated as $1- R^{m}$ \cite{May1991}, where $R$ is the basic 
reproduction number. To summarize, the average probability $q_{ij}$ of the transmission 
of a disease from $i$ to $j$ can be expressed as
\begin{align}
q_{ij} & = \sum_{m_{ij}} Q(m_{ij}) \left(  1- R^{m_{ij}} \right) \nonumber \\
	&= 1-e^{\lambda_{ij} \left( R^{-1}-1 \right)}.
\label{qij}
\end{align}

We then define $u_{ij}$ as the probability that node $j$ infected by movers from a connected 
neighbor $i$ causes a global epidemic. The probability $u_{ij}$ can be expressed 
as the product of the probability $q_{ij}$ and the probability that node $i$ is 
infected when node $j$ is absent. Therefore, the MP equations based on a local 
tree-like structure for $u_{ij}$ are defined as \cite{Karrer2014}
\begin{align}
u_{ij} = \left[ 1-e^{\lambda_{ij} \left( R^{-1}-1 \right)} \right]
\left[ 1-\prod_{k \in \partial j \setminus i} (1-u_{jk}) \right],
\label{uij}
\end{align}
where $k \in \partial j \setminus i$ represents a set of neighbors of node $j$ excluding 
node $i$. For a given metapopulation network, we can obtain $u_{ij}$ based on the numerical 
iteration of Eq.~\ref{uij}. After determining the fixed points of $u_{ij}$, we 
can obtain the probability $P_i$ of epidemic outbreaks for seed subpopulation $i$. 
A condition for a global epidemic outbreak for seed $i$ is that 
the global outbreak occurs through at least one outgoing link of node $i$.
Therefore, we arrive at the following expression for $P_i$:
\begin{align}
P_i &=  1- \prod_{j \in \partial i} (1-u_{ij}),
\label{Pi}
\end{align}
where $j \in \partial i$ represents a set of neighbors of node $i$.

We attempt to estimate $P_i$ for two extreme cases to assess the relative 
importance of various parameters, e.g., population abundance, number of 
connections, and mobility rate. Near the epidemic threshold meaning that $u_{ij}$ is small,
\begin{align}
P_i & \approx \sum_{j \in \partial i} u_{ij}.
\end{align}
For highly infectious diseases, i.e., $R \gg 1$, the degree of seed node 
affects the probability of epidemics, specifically $P_i \sim k_i$. 
It means that the network property of each node is important to assess the influence 
of highly contagious diseases. On the other hand, if $R \approx 1$, the probability 
$P_i$ is determined by the total number of outgoing infectious movers from 
$i$, such that $P_i \sim \sum_j \lambda_{ij}$. In this case, the mobility 
rate and population $n_i$ become more important.


We compute the expected size $S_i$ when global epidemic occurs with 
subpopulation $i$ (origin). We define $v_{ij}$ as the 
probability that subpopulation $i$ is infected via connected subpopulation $j$. 
The MP equations for $v_{ij}$ on a local tree-like structure are defined as
\begin{align}
v_{ij} = \left[ 1-e^{\lambda_{ji} \left( R^{-1}-1 \right)} \right]
\left[ 1-\prod_{k \in \partial j \setminus i} (1-v_{jk}) \right],
\label{vij}
\end{align}
where $k \in \partial j \setminus i$ represents a set of neighbors of node $j$ excluding 
node $i$. The direction of $v_{ij}$ is opposite to $u_{ij}$. 
When $\lambda_{ij}$ is symmetric; i.e., $\lambda_{ij}=\lambda_{ji}$, two probabilities, 
$v_{ij}$ and $u_{ij}$, become identical. Once we obtain $v_{ij}$ by solving Eq.~\ref{vij}, we 
can estimate the final size $S_i$ of epidemic outbreaks when a global epidemic occurs with 
seed subpopulation $i$:
\begin{align}
S_i &=  \frac{1}{N} \left[ 1+ \sum_{\substack{j=0 \\ j\neq i}}^N \left( 1- \prod_{k \in \partial j} (1-v_{jk}) \right) \right]\nonumber \\
&\approx \frac{1}{N} \sum_{j=1}^N \left[ 1- \prod_{k \in \partial j} (1-v_{jk}) \right],
\label{Si}
\end{align}
where $k \in \partial j$ represents a set of neighbors of node $j$.
It implies that when a global epidemic occurs, the expected final size does 
not depend much on the topological location of the seed subpopulation.


The influence $G_i$ of seed node $i$ can be obtained from the product of
the probability and the size of a global epidemic when it occurs:
\begin{align}
G_i \approx P_i S_i.
\label{Gi}
\end{align}
To recapitulate, we have determined $P_i$ and $S_i$ using the proposed
MP equations and obtained the influence $G_i$ of each node. The influence $G_i$
of node is mainly determined by the probability $P_i$ of epidemic outbreaks 
because $S_i$ is insensitive to the seed location.

\section{Results}

\begin{figure*}[t]
\includegraphics[width=\linewidth]{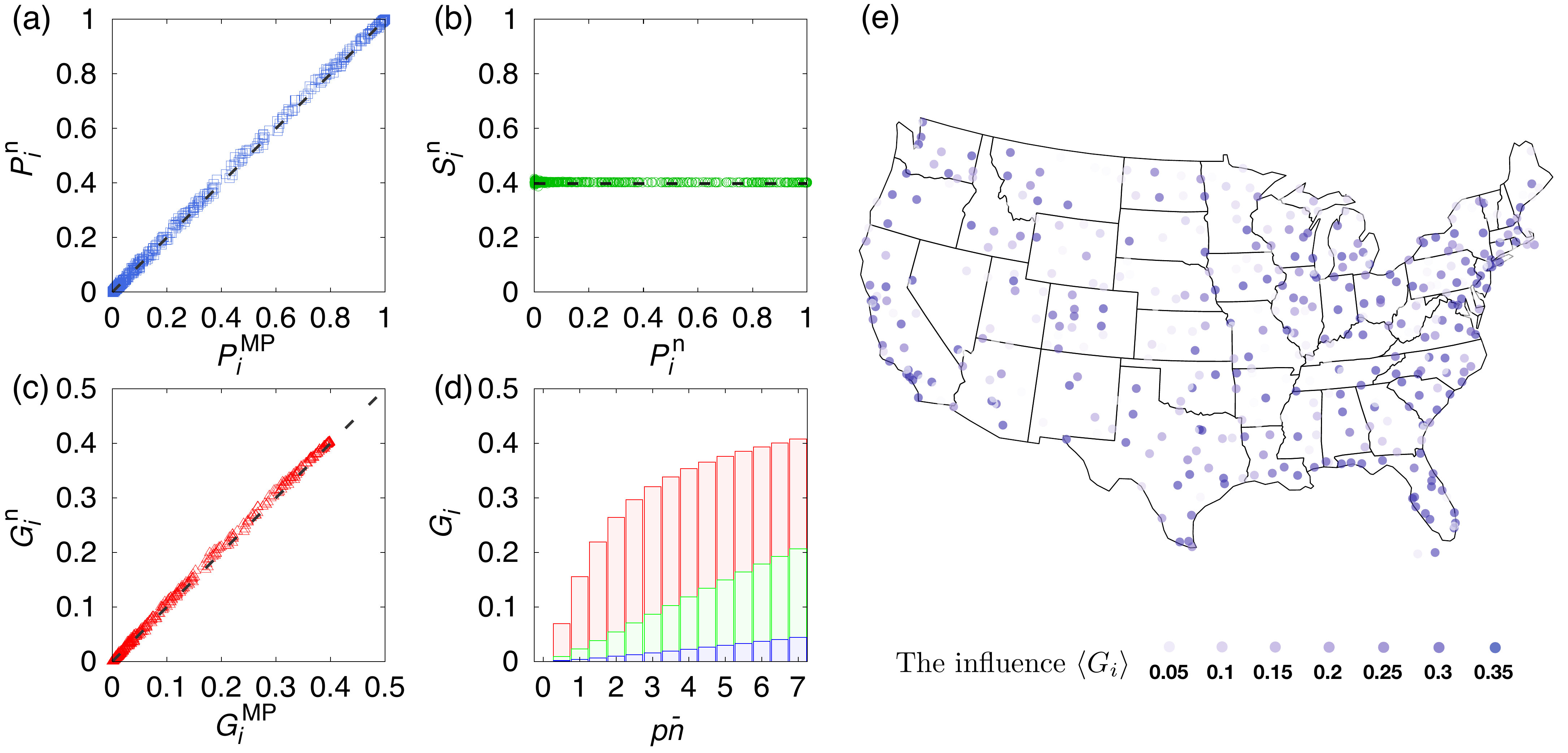}
\caption{
Influence of subpopulations of the air transportation network. 
(a) The comparison between the probability $P_i^{\mathrm{n}}$ of epidemic obtained by numerical 
simulations and the probability $P_i^{\mathrm{MP}}$ predicted using the MP theory.
(b) The size $S_i^{\mathrm{n}}$ of epidemics once global epidemic 
occurs becomes insensitive to the initial seed choice while $P_i^{\mathrm{n}}$ strongly 
depends on the seed location. The dashed line shows the size predicted 
using the theory. 
(c) The scatter plot of the influence for numerical 
simulations $G_i^{\mathrm{n}}$ and the MP theory $G_i^{\mathrm{MP}}$. 
The influence of subpopulations in the metapopulation epidemic model can be accurately 
predicted using the MP theory. $R=1.5$, $\gamma\tau=1/14$, $p \tau=6.29 \times 10^{-6}$,
and $\tau=1$ are used for (a-c). 
(d) Examples of $G_i$ for three different cities: Allentown/Bethlehem/Easton (red),
Philadelphia (green), Devils Lake (blue)
(e) The average $\langle G_i \rangle$ of the influence 
for various migration rate $p$ indicates the influence of each subpopulation. 
The influence map of cities in the air transportation network 
based on $\langle G_i \rangle$.
}
\label{fig2}
\end{figure*}

\subsection{Air Transportation Network}

An air transportation network sparsely connecting cities via airlines is a prominent 
example of metapopulation models. In this study, we construct a metapopulation network 
based on the US air transportation data from the origin to destination airport market data 
gathered from Bureau of transportation statistics \cite{BTS}. 
We considered the air transportation network as a weighted and directional network.
Airports (cities) in the network correspond to nodes, and airlines correspond to links.
The air transportation network includes 1220 nodes (airports) and 20,756 outgoing links (traffic). 
The weight and directionality of links are defined by the total number of passengers ($T_{ij}$) 
between two airports during $2019$. 
The population of each city was estimated to be proportional to the total number of 
passengers leaving the city, which is expressed as follows:
\begin{align}
n_i = \alpha \sum_{j \in \partial i} T_{ij},
\end{align}
where $\alpha$ is a constant. For simplicity, in this study, we set $\alpha=1$.
The average population over all the cities is about $1.1 \times 10^6$. There is no significant 
difference in the results as long as the total population is large enough because 
metapopulation modeling assumes a well-mixed internal structure.
We only used nodes whose total population (the total number of outgoing passengers) is larger 
than 100 for reliable predictions.
Each individual moves to one of neighbor cities with migration rate $p$.
Thus, the migration probability from $i$ to $j$ for each individual in node $i$ 
becomes $p T_{ij} \tau/n_i$. 
We performed numerical simulations of the spread of infectious diseases using the 
aforementioned setting of the air transportation network. Specifically, by varying seed 
node $i$, we numerically measured $P_i$, $S_i$, and $G_i$ for $R=1.5$, $\gamma \tau=1/14$, and 
$p \tau=6.29 \times 10^{-6}$, and $\tau=1$. We confirmed that our results are robust for 
a wide range of parameters.

We validated the MP theory for the influence of each node on the air transportation 
network. 
Each infectious individual sustains infective approximately for an 
average time of $1/\gamma$. 
Therefore, the average number $\lambda_{ij}$ of infectious 
movers from $i$ to $j$ can be inferred as
\begin{align}
\lambda_{ij} &= \frac{r p T_{ij}}{\gamma}~,
\end{align}
where $r$ is the average fraction of infected individuals in nodes in the epidemic 
phase. The average fraction $r$ of infected individuals in node can be obtained in 
a well-mixed population by solving the self-consistency equation $r= 1-e^{-r R}$,
where $R$ is the basic reproduction number \cite{May1991}.
Substituting $\lambda_{ij}$ into the MP equations, we can obtain the probability of epidemic 
outbreak ($P_{i}^{\mathrm{MP}}$), the size of epidemic outbreak ($S_{i}^{\mathrm{MP}}$), 
and the influence of subpopulations ($G_{i}^{\mathrm{MP}}$).


Our analytical predictions using the MP equations well agree with the numerical simulations. 
First, the scatter plot of $P_i$ shows the relationship between the theoretical 
and numerical results [Fig.~2(a)]. Specifically, the probabilities $P_i$ of epidemic outbreak 
initiated by an infectious individual in subpopulation $i$ calculated based on the MP theory 
$P_i^{\mathrm{MP}}$ and numerical simulations $P_i^{\mathrm{n}}$ are shown in Fig.~\ref{fig2}(a).
Each point in Fig.~2(a) represents the comparison between the MP prediction and numerical 
results, calculated from a different seed. The result shows that the MP approach allows us 
to accurately determine the probability of epidemic outbreaks for different initiators.
Second, we found that once an epidemic outbreaks globally, its size becomes insensitive 
to the location of the seed subpopulation. We compare the size $S_i^{\mathrm{n}}$ and
the probability $P_i^\mathrm{n}$ for various seed nodes in Fig.~\ref{fig2}(b). While the 
probability $P_i$ of epidemic outbreak significantly depends on the location of seed 
selected, its size $S_i$, once an outbreak occurs, is insensitive to different seed nodes.
as predicted using the MP theory.
Moreover, we confirm that the numerical results well agree 
with the size $S_i^{\mathrm{MP}}$ predicted using the MP theory indicated by the horizontal 
dashed line in Fig.~2(b). Finally, the numerical results $G_i^\mathrm{n}$ with regard to 
the influence of each node agree with the theory $G_i^{\mathrm{MP}}$ [Fig.~2(c)]. 
The variability of $G_i$ is large depending on the seed subpopulation, implying that the 
difference in influence is significant for each subpopulation.
Figure~2(c) shows that $G_i$ with such large variability of $G_i$ can be accurately predicted using the MP theory.

To analyze the general influence $\langle G_i \rangle$ of each city, we measure
$G_i^{\mathrm{MP}}$, which has been averaged over various migration rate $p$. As 
shown in Fig. 2(d), the influence $G_i$ for various values of the rate $p$ 
was calculated using the MP equation. Then, the average $\langle G_i \rangle$ of each node  
was obtained, and it was used as the average influence index of node $i$ in the metapopulation model. 
Figure 2(d) shows examples of $G_i$ for three different cities 
i.e.,	Allentown/Bethlehem/Easton (red), Philadelphia (green), and Devils Lake (blue)
as a function of $p \bar{n}$. Three cities show different $G_i$ for various values of $p \bar{n}$.
An influence map can be obtained using the influence index 
$\langle G_i \rangle$. The influence of each city on the metapopulation epidemic model 
in the US air transport network is depicted in Fig.~2(e). The higher the influence, the darker 
and bluer the symbols are on the map. The top 10 influential nodes (airports) are listed
in Table 1. We also indicated in Table 1 their relative influence $\langle G_i \rangle$,
degree \cite{Albert2000}, betweenness centrality \cite{Freeman1977}, and node 
strength \cite{Barrat2004}. The relative influence is measured by the ratio of $\langle G_i \rangle$
to the influence of the top node, $G_\mathrm{o}$. 
Our analysis comprising mathematical modeling, the MP theory, and real-world data allows 
us to systematically predict dangerous cities with regard to epidemic spreading.

\begin{table}
\begin{tabular}{|c|c|c|c|c|}
\hline
\multirow{2}{*}{airport} &\multirow{2}{*}{$\langle G_i \rangle / G_\mathrm{o} $ }&\multirow{2}{*}{Degree} & Betweenness & Node \\
 ~&~& ~&centrality & strength\\
\hline
ATL&1&200&0.027&$4.7\times10^7$ \\
LAX&0.9990&176&0.027&$3.0\times10^7$ \\
ORD&0.9986&216&0.057&$3.4\times10^7$ \\
DEN&0.9977&219&0.066&$3.2\times10^7$ \\
DFW&0.9962&209&0.047&$3.1\times10^7$ \\
LAS&0.9940&175&0.029&$2.3\times10^7$ \\
SFO&0.9932&143&0.016&$2.0\times10^7$ \\
MCO&0.9923&137&0.011&$2.1\times10^7$ \\
PHX&0.9919&159&0.019&$2.1\times10^7$ \\
SEA&0.9913&135&0.050&$2.2\times10^7$ \\
\hline
\end{tabular}
\caption{The list of the top 10 influential nodes (airports) of the U.S. air transportation network. 
The airports are listed in the order of the average influence $\langle G_i \rangle$ in the range 
of $0 < p\bar{n} < 0.385$.
}
\end{table}



\subsection{Random Networks}

To confirm the generality of our theory, we check our theory on random, Erd\"os-R\'enyi 
(ER), and scale-free (SF) networks. We use ER networks with $N=10^4$ and $\langle k \rangle =4$ 
and SF networks with $N=10^4$ and degree exponent $\gamma_k=2.5$. To generate random SF networks, 
we implement a static SF model \cite{Goh2001}.
We calculated the epidemic probability and size in these networks and compared the numerical 
results with the predictions of the MP theory. To compute the MP theory on random networks, 
we infer $\lambda_{ij}$ because the number of passengers ($T_{ij}$) is not given a priori 
unlike the air transportation network. For simplicity, we assume that all individuals move with 
the same migration rate $p$ \cite{Colizza2006,Colizza2008}. 
Individuals in a subpopulation exhibit the same probability to move to one of its 
neighbors as $p \tau/k$ when the probability that an individual leaves a subpopulation 
is homogeneous and independent of its degree $k$. In the steady state, the number 
of individuals arriving at a node with degree $k$ is proportional to degree $k$. 
Then, the average population of a node with degree $k$ at a steady state is defined by 
$n_k = \bar{n}k/{\langle k \rangle}$, where $\bar{n}$ is the average 
population of all the nodes. The average number of movers 
to a neighbor node is defined by the product of the individual 
migration rate to one neighbor, $p/k$, and the total population of a subpopulation, $n_k$. 
According to the aforementioned relationships, the average number  of 
infectious movers $\lambda$ is given by \cite{Colizza2008}
\begin{align}
\lambda= \frac{r \bar{n} p}{\gamma \langle k \rangle},
\label{lambda}
\end{align}
where $\bar{n}$ is the average population of all the nodes.
Thus, we obtain the probability, size, and influence of epidemic outbreaks 
for each node using the MP theory with Eqs. \ref{uij}–\ref{Gi}.


We further investigate the average size of epidemics for metapopulation models for the ensemble
of random networks with a given degree distribution. We introduce annealed approximation and arrive 
at the following self-consistency equation:
\begin{align}
u&= \left[ 1-e^{\lambda (R^{-1}-1)} \right] \left[ 1- \sum_{k=1}^{\infty} \frac{k P(k)}{\langle k \rangle} (1-u)^{k-1} \right],
\label{u_random}
\end{align} 
where $u$ is the probability that a node has been infected by a neighbor node
and $P(k)$ is the degree of distribution.
Once Eq.~\ref{u_random} is solved using numerical iterations, we can obtain the 
average size $S$ of epidemics as
\begin{align}
S= 1- \sum_{k=0}^{\infty} P(k) (1-u)^k. 
\label{S_random}
\end{align}

\begin{figure}
\includegraphics[width=\linewidth]{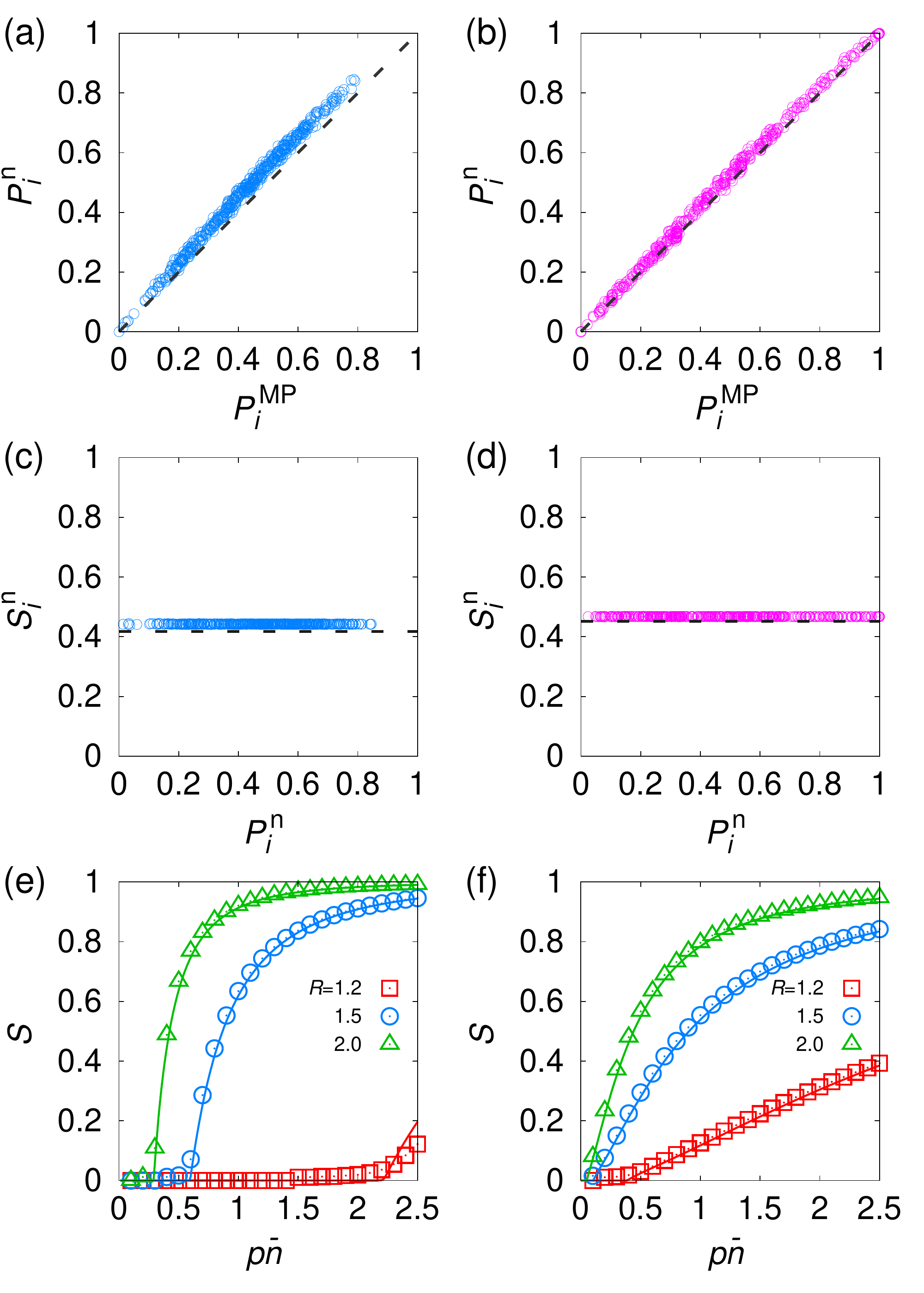}
\caption{
Influence of subpopulations on random networks. The comparison between 
the probability $P_i^{\mathrm{n}}$ obtained using numerical simulations and $P_i^{\mathrm{MP}}$ 
obtained using the MP theory is depicted by for (a) ER and (b) SF networks.
The size $S_i^{\mathrm{n}}$ of epidemics once global epidemic occurs 
becomes insensitive to the initial seed for (c) ER and 
(d) SF networks. The average size 
predicted based on the annealed approximation are given by 
for (e) ER and (f) SF networks.
We use ER networks with $N=10^4$, $\langle k \rangle =4$, and $\bar{n}=10^4$ and SF networks 
with $N=10^4$, $\gamma_k=2.5$ and $\langle k \rangle =4$, and $\bar{n}=10^4$.  
$R=1.5$, $\gamma\tau=0.1$, $p \tau=0.8 \times 10^{-4}$, and $\tau=1$ 
	are used for (a-d).
Simulation results are averaged over $5000$ different realizations.
}
\label{fig3}
\end{figure}

In addition, we can predict the epidemic threshold $p_c$ of the mobility parameter.
The trivial solution for Eq.~\ref{u_random} is $u = 0$. This solution corresponds 
to the disease-free phase. As $p$ increases, another solution can appear. Such 
threshold point can be obtained by applying a linear stability analysis near
$u = 0$. Imposing the condition for the epidemic threshold, we arrive at
\begin{align}
1&= \left[ 1-e^{\lambda_c (R^{-1}-1)} \right] \left[ 1- \frac{\langle k^2 \rangle - \langle k \rangle}{\langle k\rangle} \right].
\end{align}
As $\lambda= \frac{r \bar{n} p}{\gamma \langle k \rangle}$, the epidemic threshold 
can be derived as
\begin{align}
p_c \bar{n} & = \frac{\gamma R \langle k \rangle}{r(1-R)} \log \left( 1- \frac{\langle k \rangle}{\langle k^2 \rangle - \langle k \rangle} \right). 
\end{align}
The threshold $p_c$ predicted using the MP theory is more accurate compared with the previous 
mean-field approximation \cite{Colizza2007}.

In Figs. 3(a–d), we compare numerical results and theoretical predictions for epidemic 
probability and size in ER and SF networks. We confirm that the proposed MP approach enables 
accurate identification of influential subpopulations in metapopulation epidemic models.
The probabilities of epidemic outbreak initiated by an infectious seed in subpopulation $i$ 
obtained using the MP theory $P_i^{\mathrm{MP}}$ and numerical simulations $P_i^{\mathrm{n}}$ 
are depicted in Fig.~\ref{fig3}(a) for ER and Fig.~\ref{fig3}(b) for SF. The probability 
$P_i^{\mathrm{MP}}$ accurately predicted based on the MP theory agrees with the 
simulation results. Moreover, we examined the average size ($S_i^{\mathrm{n}}$) of an 
epidemic outbreak when it occurs with respect to $P_i^{\mathrm{n}}$ defined in Fig.~3(c) for 
ER and Fig.~3(d) for SF. As we predicted based on the MP theory, the probability depends on 
the location of the initial seed, but the size is insensitive. In addition, the epidemic 
size $S$ predicted based on an annealed approximation
are shown in Fig.~3(e) for ER and Fig.~3(f) for SF. 
We confirm that the epidemic size and threshold obtained above agree well with 
the simulation results for ER and SF networks with various $\beta$ and $\gamma$.
Although our theory predicts numerical results well, there is still a discrepancy 
between them. The reason for the discrepancy is that we applied several approximations 
that are not completely exact when calculating $q_{ij}$. For instance, we neglect 
the effects of the number of multiple infectious seeds for estimating $r$ and 
the fluctuations of populations of each node due to the migration of individuals.

\subsection{Comparison with Other Centralities}

We compare the MP theory and conventional centralities for their ability to identify 
influential nodes in metapopulation epidemic models. We use the following centralities:
high degree \cite{Albert2000}, $k$-core \cite{Dorogovtsev2006}, betweenness 
centrality \cite{Freeman1977,Goh2001}, and node strength \cite{Barrat2004}.
The degree $k$ analyzes the local importance of a node, $k$-core $k^{\mathrm{co}}$ represents the topological 
importance of the node, betweenness centrality $C$ represents the centrality based on the shortest 
path on a network, and node strength $w$ represents the sum of outgoing link weights. 
We computed the centralities by using \textrm{NetworkX} package.
In addition, we treat networks as bi-directed and unweighted 
when we calculate degree, $k$-core index, and betweenness centrality. 
To compare the capability of centralities to predict the influence of spreading 
processes, we calculated the Pearson correlation coefficient $r$ between 
numerical results of the influence $G_i^\mathrm{n}$ and centralities for the US 
airline, ER, and SF networks.
Note that node strength $w$ is linearly correlated to degree $k$ for ER and SF networks
in our model setting \cite{Colizza2007}, so that $r$ for the two centralities are identical.

\begin{figure}[!]
\includegraphics[width=\linewidth]{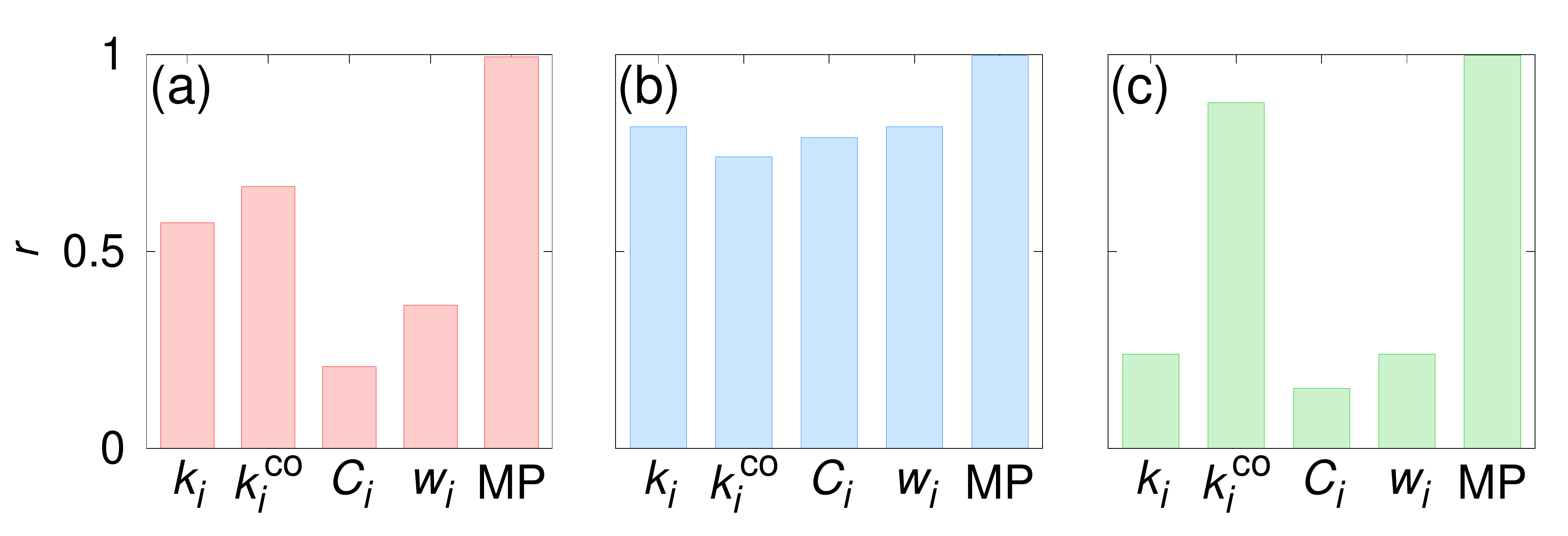}
\caption{
Comparison of the influence of subpopulations between other centralities 
for (a) air transportation, (b) ER, and (c) SF networks. 
Pearson correlation coefficients between numerical results of 
the influence ($G_i^n$) and other centralities, degree $k$, 
 $k$-core index $k^{\mathrm{co}}$, betweenness centrality $C$, node strength $w$, 
 and MP theory $MP$ are presented in the figure. 
 We used the same parameters of Fig.~\ref{fig2}(a-c) for air transportation network and 
 Fig.~\ref{fig3}(a-d) for ER and SF networks for the calculation of $r$.
The MP theory outperforms the existing centralities to identify influential 
subpopulations in metapopulation models. 
}
\label{fig4}
\end{figure}

Figure 4 shows correlation coefficients for degree $k$, $k$-core index $k^{\mathrm{co}}$, betweenness 
centrality $C$, node strength $w$, and the MP theory ${\mathrm{MP}}$. The correlation between the MP 
theory and numerical results exceeds $0.99$ for all the tested networks, indicating that the predicted 
values accurately agree with the calculated values. However, the other centralities exhibit 
significantly lower correlation coefficients compared with the MP theory. 
Moreover, the conventional centralities do not produce reliable predictions as the correlation 
coefficient for $k$, $k^{\mathrm{co}}$, $C$, $w$ exhibit different tendencies for different 
network structures. This result implies that centralities based on network topology are 
insufficient to predict the influence of subpopulations in metapopulation models. The proposed 
MP method considers the network topology and dynamical processes and outperforms the existing 
centralities in terms of accuracy and reliability for determining influential subpopulations 
in metapopulation models.

\section{Conclusion}

In this study, we derive the MP theory for the metapopulation epidemic model and 
propose a method to identify influential subpopulations. We show how to leverage 
these analyses based on real-world data and how to determine the most dangerous 
cities as an initial seed of potential epidemics. We validate the theory 
based on extensive numerical simulations on empirical and random networks with 
a broad range of parameter sets. As a result, it is confirmed that using the proposed 
theory, we can accurately and robustly predict important subpopulations in metapopulation 
models. In addition, we show that the information of the network topology is insufficient 
to accurately determine important nodes in metapopulation models. We also found that the 
location of initial spreaders significantly affects the probability of an epidemic 
outbreak but not the average size of an epidemic outbreak once a large-scale
epidemic occurs. In this study, the method of identifying influential subpopulations 
can significantly impact control and intervention policies defined for global epidemics.
Further study is required to examine the effect of the response and vaccination 
with different subpopulations on the spread of epidemics on metapopulation 
modeling \cite{Maier2020,Zhong2021,Tanaka2014}.

\acknowledgements

This work was supported by the National Research Foundation of Korea (NRF) grants 
funded by the Korean Government (MSIT) (No. 2020R1I1A3068803).

\bibliography{./spreaders}

\end{document}